\documentclass[prb,twocolumn,floatfix,showpacs,preprintnumbers,superscriptaddress,amsmath,amssymb,eqsecnum]{revtex4} 

\usepackage{graphicx}

\begin{document}

\title{Persistent Spin Currents in Helimagnets}

\author{Jan Heurich}
\author{J\"urgen K\"onig} 
\affiliation{Institut f\"ur Theoretische Festk\"orperphysik, Universit\"at Karlsruhe, 
76128 Karlsruhe, Germany\\}
\author{A. H. MacDonald}
\affiliation{Department of Physics, University of Texas at Austin, Austin, 
TX 78712}

\date{\today}

\begin{abstract}

We demonstrate that weak external magnetic fields generate dissipationless 
spin currents in the ground state of systems with spiral magnetic order.
Our conclusions are based on phenomenological considerations and on 
microscopic mean-field theory calculations for an illustrative toy model.
We speculate on possible applications of this effect in spintronic devices.  

\end{abstract}

% 75.10.Lp Magnetic Properties and Materials: Band and itinerant models 
% 72.25.Ba Spin polarized transport in metals
% 74.20.-z Superconductivity: Theories and models of superconducting state 
% 72.15.Nj Electronic transport in condensed matter: Collective modes 

\maketitle

\section{Introduction}

Collective transport effects in ordered many-fermion and many-boson systems 
include some of the most dramatic and profound phenomena that occur 
in condensed matter physics.  For example,  
the (practically) dissipationless transport of electrical charge by
Cooper pairs\cite{bcs} in superconductors and superfluidity in $^{4}$He and 
$^{3}$He have been important topics through 
most of the field's history.  New instances of this general class of phenomenon 
continue to arise and create interest.  
One recent case is collective charge transport in  
double-layer quantum Hall systems, in which spontaneous inter-layer phase 
coherence leads to a strongly enhanced zero-bias tunnel current from one
layer to the other.\cite{qheold,jordan,qheeisen,tuntheory}
Closely related issues connected with the possibility 
of superfluidity due to excitonic Bose condensation 
in electron-hole double-layer systems\cite{eh_pair,exc} 
also continue to attract attention\cite{excitonbec} and inspire 
experimental activity.

\begin{figure}
\centerline{\includegraphics[width=0.9\columnwidth]{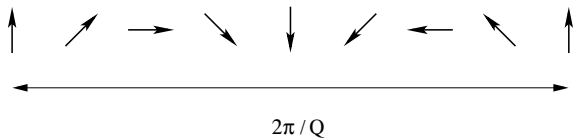}}
\caption{State with spiral magnetic order in the $\hat{\bf x}$-$\hat{\bf y}$ 
  plane characterized by the wave vector $Q$.
  The arrows represent the local spin density of the itinerant-electron 
  system.}
\label{fig:one}
\end{figure}

Two of us\cite{us} have recently proposed the possibility of 
realizing nearly dissipationless collective spin currents in easy-plane
thin-film ferromagnets. Although the proposed phenomenon has some formal 
relationship to superconductivity and superfluidity, it depends in 
part on a symmetry (magnetization orientation invariance within an easy 
plane) that can only be realized approximately.  More importantly, unlike the 
case of superconductors which can easily be biased by a current source, 
strategies for driving a real thin-film ferromagnet into the metastable 
spin-current state present experimental and materials challenges that have 
not yet been overcome.
For ferromagnets, the meta-stable states that have a non-zero spin current 
are spiral states, as sketched in Fig.~\ref{fig:one}.

In Ref.~\onlinecite{us} we demonstrated that these 
states are metastable only when the system has a non-zero magnetic anisotropy
that is accurately uniaxial with an easy plane, and proposed an experimental
strategy to generate and detect macroscopic 
dissipationless spin currents in thin-film ferromagnets. 
Other related recent work has addressed persistent spin currents
in rings that experience an inhomogeneous magnetic field.\cite{loss,gao,kopietz}
In this paper we consider the case in which the {\em ground} state of the system
has spiral order, the case of {\em helimagnets}. 
As mentioned already in Ref.~\onlinecite{us}, the persistent
spin current of a spiral state is proportional to the derivative of its 
energy with respect to the state's winding wavevector $\bf Q$, a quantity which
necessarily vanishes in the ground state.  Like ferromagnets, helimagnets do
not normally carry a spin current in their ground state.  However, as we demonstrate 
in this paper, in the helimagnet case 
a persistent ground state spin current is induced by an external magnetic
field.

Throughout this paper we consider states in which the spin density 
$\langle{\bf s}({\bf r})\rangle$ lies in the $\hat {\bf x}$-$\hat{\bf y}$ 
plane.  The stability of these states will, in general, require some easy-plane
anisotropy that could have magnetostatic or magnetocrystalline origin.
The presence of such a magnetic anisotropy energy term is implicitly 
assumed throughout the paper.  
In the first part of the paper, we set the applied magnetic field to zero and consider 
states with simple spiral order, illustrated schematically in Fig.~\ref{fig:one}.
Our objective in this section is to demonstrate by an explicit microscopic calculation 
the property that the spiral ground states of helimagnets, unlike the 
spiral metastable excited states of ferromagnets, do not 
carry persistent spin currents.  The order parameter of a spiral state is 
\begin{equation}
   \langle{\bf s}({\bf r})\rangle_{\bf Q} = s_{\bf Q} [\cos({\bf Q} \cdot 
        {\bf r})
        \hat {\bf x} \mp \sin({\bf Q} \cdot {\bf r}) \hat {\bf y}] \, .
\end{equation}
and it is characterized by a wave vector $\bf Q$, by the  
magnitude $s_{\bf Q}$ of the order parameter, and by a chirality.
For a single-band lattice model $0 \le s_{\bf Q} \le 0.5 
n_{\rm e}$, where $n_{\rm e}$ is the electron density per atomic site.
As $\bf Q$ is varied, the change of the order parameter's spatial dependence 
will cause a change in the magnetic condensation energy and in the magnitude 
of the order parameter.
We have argued previously\cite{us} that these systems carry a dissipationless  
spin current, {\em i.e.}, a current with equal magnitude and 
opposite directions for up and down spins, that is related to the dependence of the 
total energy density $\epsilon$ on $\bf Q$ by\cite{us},
\begin{equation}
   {\bf j}_{\uparrow} = 
        {e \over \hbar} {\partial \epsilon ({\bf Q}) \over \partial {\bf Q}} 
        = - {\bf j}_{\downarrow} \, .
\label{eq:curdens}
\end{equation}
Note that the spin quantization axis is along the $\hat {\bf z}$-direction here.  
All spin currents discussed in this paper are polarized along the direction
perpendicular to the plane defined by the spiral order parameter; we therefore
do not explicitly indicate the tensor character of spin currents. 
Eq.~(\ref{eq:curdens}) implies that it is not the presence  
of spiral magnetic configurations, but rather the change of total energy with varying
spiral wave vector $\bf Q$ which leads to spin currents \cite{slonczewski}.
If the energy has a local minimum at a finite wave vector $\bf Q^\star$
(as in a helimagnet) then 
this state will, unlike spiral states in a system with a ferromagnetic 
ground state, not support a persistent spin current.  We examine this 
property from a microscopic point of view in Section II. 
In Section III we discuss persistent spin currents using
the picture of ``Cooper pairs'' familiar from the
Bardeen-Cooper-Schrieffer (BCS)
mean-field theory of superconductivity, which provides an alternative 
way of understanding when and why spin supercurrents
appear for a given magnetization configuration. 

With this background, we turn our attention in Section IV to a 
new strategy for generating
persistent spin currents.  We point out that persistent spin currents can be 
generated in systems with spiral magnetic order simply by applying a
magnetic field oriented in the plane of spiral order, 
the $\hat {\bf x}$-$\hat{\bf y}$ plane
in our notation.  The external magnetic field creates a competition between spiral order
with wave vector $\bf Q^\star$ and homogeneous magnetization along the field direction,
leading to magnetic order described by a soliton lattice with wave vector 
$|{\bf Q}| < |{\bf Q^\star}|$.  We demonstrate that soliton-lattice 
states carry persistent spin 
currents, whose amplitude is controlled by the strength of the applied 
magnetic field.  Finally, we conclude in Section V with some brief speculations
on possible applications in spintronics for these field-induced spin currents. 

\section{Microscopic Model for Persistent Spin Currents in Helimagnets}

\subsection{Hartree-Fock theory for spiral magnetic order}

In this section we try to shed additional light on the relationship between
persistent spin currents and the dependence of energy on the spiral wavevector by
carrying out explicit calculations for a simple toy model of an itinerant-electron 
system with magnetic order.  Similar spiral state models have been presented 
previously\cite{capellmann,auerbach}; the following brief description is included
for completeness and intended to establish notation for the following discussion.

We consider a system of fermions on a lattice with single-particle band energy  
$\epsilon_{\bf k}$ and delta-function repulsive particle-particle interaction 
$U \delta ({\bf r}_i - {\bf r}_j)$, which we treat in a mean-field approximation. 
The unrestricted Hartree-Fock mean-field Hamiltonian for the spiral ordered state with 
wave vector ${\bf Q}$ is
\begin{widetext}
\begin{equation}
   {\cal H}^{\rm HF} = {Vh^2\over U} +
        \sum_{\bf k}
        \left( c_{{\bf k} + {\bf Q}/2,\uparrow}^\dagger \, \, 
                c_{{\bf k} - {\bf Q}/2,\downarrow}^\dagger
        \right)
        \left(
        \begin{array}{cc}
                \epsilon_{{\bf k} + {\bf Q}/2} & -h \\
                -h & \epsilon_{{\bf k} - {\bf Q}/2}
        \end{array} \right)
        \left(
        \begin{array}{c}
                c_{{\bf k} + {\bf Q}/2,\uparrow}\\
                c_{{\bf k} - {\bf Q}/2,\downarrow}
        \end{array} \right)
\label{eq:mft}
\end{equation}
\end{widetext}
\noindent
where $h = U s_{\bf Q}$ and
$s_{\bf Q} = (1/V){\sum_{\bf k}} \langle 
c^\dagger_{{\bf k}+{\bf Q}/2,\uparrow} c_{{\bf k}-{\bf Q}/2,\downarrow} 
\rangle$.
The off-diagonal terms in Eq.~(\ref{eq:mft}) couple electronic states with 
opposite spin and Bloch wave-vector difference ${\bf Q}$.
(The wavevectors above are understood to be reduced to the first Brillouin-zone of
the lattice.) To diagonalize the Hamiltonian we employ the transformation
\begin{equation}
  \left(
    \begin{array}{c}
      a_{{\bf k},+}\\
      a_{{\bf k},-}
    \end{array} \right)
  =
  \left(
    \begin{array}{cc}
      \cos \Theta_{\bf k} & -\sin \Theta_{\bf k} \\ 
      \sin \Theta_{\bf k} & \cos \Theta_{\bf k} 
    \end{array} \right)
  \left(
    \begin{array}{c}
      c_{{\bf k} + {\bf Q}/2,\uparrow}\\
      c_{{\bf k} - {\bf Q}/2,\downarrow}
    \end{array} \right)
\end{equation}
with $\tan 2\Theta_{\bf k} = 2h / [ \epsilon_{{\bf k} + {\bf Q}/2} -
\epsilon_{{\bf k} - {\bf Q}/2} ]$
and $0\le \Theta_{\bf k} < \pi/2$ to arrive at
${\cal H}^{\rm HF} = (Vh^2 / U) + \sum_{{\bf k},\pm} E_{\bf k}^\pm
a^{\dagger}_{{\bf k},\pm} a_{{\bf k},\pm}$.
The eigenenergies of the ordered-state quasiparticles are given (for $h\ge 0$)
by 
\begin{equation}
   E_{\bf k}^\pm = { \epsilon_{{\bf k} + {\bf Q}/2} +
        \epsilon_{{\bf k} - {\bf Q}/2} \pm 
        \sqrt{(\epsilon_{{\bf k} + {\bf Q}/2} -
        \epsilon_{{\bf k} - {\bf Q}/2})^2 + 4 h^2} \over 2}.
\label{eq:qpenergies}
\end{equation}
The effective magnetic field which helps split the quasiparticle bands 
is fixed by solving the self-consistency equation
$h = U s_{\bf Q}$.  For this simple model an explicit expression can be
given for the right hand side and we obtain 
\begin{equation}
  {U\over V}{\sum_{\bf k}}
  \frac{f(E^-_{\bf k}) - f(E^+_{\bf k})}
  {\sqrt{(\epsilon_{{\bf k} + {\bf Q}/2} 
      - \epsilon_{{\bf k} - {\bf Q}/2})^2 + 4 h^2}}  = 1 \, ,
\label{eq:selfconsistent}
\end{equation}
where $ f(E)$ is the zero-temperature Fermi function with the chemical 
potential determined by 
$(1/V)\sum_{\bf k} [f(E^-_{\bf k})+f(E^+_{\bf k})] = n_{\rm e}$. 

\subsection{Specific toy model}

The above equations are valid for an arbitrary band dispersion relation $\epsilon_{\bf k}$. 
In Ref.~\onlinecite{us} free fermions with parabolic dispersion were 
considered. For this case states with spiral magnetic order always have higher
energy than uniform magnetization (ferromagnetic) states.
In the present paper, however, we want to construct a model in which the 
minimum (mean-field theory) energy occurs for a spiral state with wave vector 
${\bf Q^\star}\neq 0$, {\it i.e.}, 
a model for which spiral magnetic order is favored over ferromagnetism.\cite{stoner}
This circumstance is achieved most simply by choosing a quasi-one
dimensional model, {\it i.e.}, by choosing 
\begin{equation}
  \epsilon_{\bf k} = {W\over 2} \left[ 1 - \cos (k_z a) \right]
\end{equation}
for $-\pi/a \le k_z \le \pi/a$ independent of $k_x$ and $k_y$, where $W$ 
defines the bandwidth.  It is also convenient to consider the case of a 
half-filled band, {\it i.e.} $a = n_{\rm e}^{-1/3}$, where $n_{\rm e}$ is 
the electron density.
We study this pedagogical toy model in order to 
illustrate the general relationship discussed earlier\cite{us}
between spiral magnetic order and spin supercurrents, not with the 
objective of modeling any specific material.  In practice helimagnetism
occurs for various reasons in several different types of materials;
see for example Ref.~\onlinecite{kawamura} and work cited therein.  
We will briefly discuss some of these materials in Sections IV
and V.

Choosing a quantization axis perpendicular to the spiral-order plane,
the spin-projected current density is given by 
\begin{equation}
  {\bf j}_\sigma = {e\over V} \sum_{\bf k} {\partial \epsilon_{\bf k} \over
    \partial (\hbar {\bf k})} 
  \langle c^\dagger_{{\bf k}\sigma} c_{{\bf k}\sigma} \rangle \, .
\label{eq:defj}
\end{equation}
The number operator average in the mean-field-theory state 
can be expressed in terms of Fermi occupation factors for the 
quasiparticles of the Hartree-Fock Hamiltonian leading to
\begin{equation}
  {\bf j}_\uparrow = {e\over V} \sum_{\bf k} 
  {\partial \epsilon_{{\bf k}+{\bf Q}/2} \over
    \partial (\hbar {\bf k})} 
  \left[ \sin^2 \Theta_{\bf k} \; f( E^-_{\bf k} ) +
    \cos^2 \Theta_{\bf k} \; f( E^+_{\bf k} ) \right] \, .
\label{eq:direct}
\end{equation}
It is then straightforward to show that
$\epsilon_{\bf k} = \epsilon_{- \bf k}$ implies that  
${\bf j}_\downarrow = -{\bf j}_\uparrow$.

An alternative expression follows from Eq.~(\ref{eq:curdens}), which was 
discussed earlier\cite{us} and is explained in more detail in Appendix~\ref{proof}.
The mean-field theory expression for the energy of a state with
spiral-wavevector ${\bf Q}$ is 
\begin{equation}
  \epsilon({\bf Q}) = {h^2\over U} + {1\over V} \sum_{\bf k}
  \left[ E^+_{\bf k} \; f( E^+_{\bf k} ) + E^-_{\bf k} \; f( E^-_{\bf k} ) \right].
\label{eq:meanfieldenergy} 
\end{equation}
For all the numerical results discussed in this paper we explicitly checked the 
equivalence of Eqs.~(\ref{eq:curdens}) and (\ref{eq:direct}).
Note that a spin current can be present even though the quasiparticle population is
in equilibrium. Elastic scattering from occupied to unoccupied quasiparticle 
states cannot provide a spin-current decay mechanism.
The spin currents are instead carried collectively, as in the case of 
dissipationless charge transport in BCS superconductors, and can decay only
by collective processes\cite{phaseslips} that allow the phase, which specifies 
the spiraling magnetization orientation, to slip.

\subsection{Numerical results}

\begin{figure}
\centerline{\includegraphics[width=0.9\columnwidth]{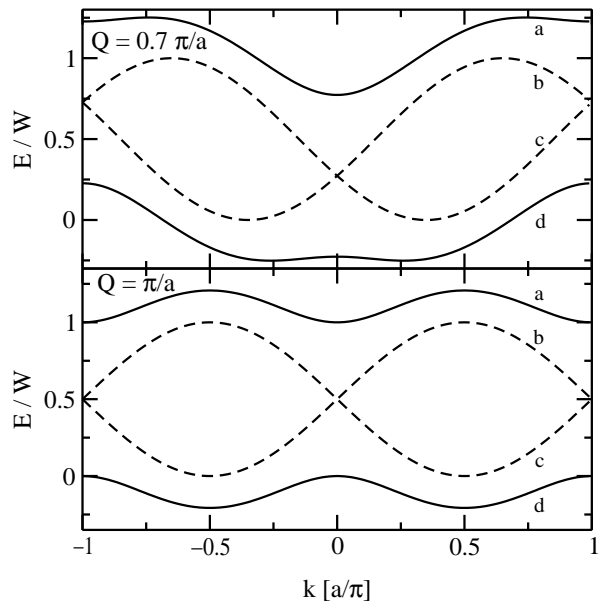}}
\caption{Quasiparticle bands $E_{k_z}^+$ (a) and $E_{k_z}^-$ (d) for two 
  different values of $Q$.
  The interaction strength $U$ is characterized by 
  $2n_{\rm e}U/(\pi W) = 1$.
  For comparison, we also show the dispersion relations $\epsilon_{k_z+Q/2}$ (b) and
  $\epsilon_{k_z-Q/2}$ (c) for zero order parameter as dashed lines.}
\label{fig:two}
\end{figure}

\begin{figure}
\centerline{\includegraphics[width=0.9\columnwidth]{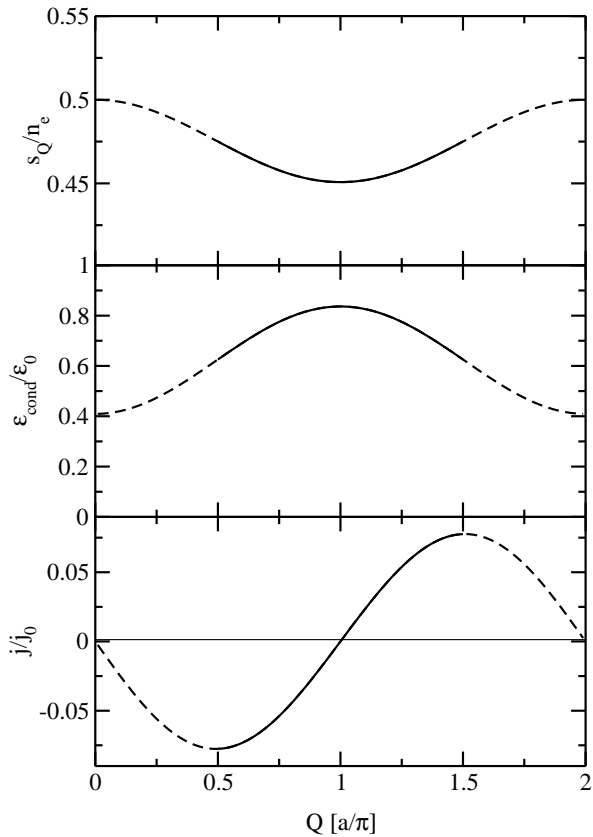}}
\caption{The order parameter $s_Q$ normalized to the electron density $n_{\rm e}$, 
the magnetic condensation-energy density $\epsilon_{\rm cond}$ normalized to 
the energy density $\epsilon_0$ of the disordered state, 
and the spin supercurrent density $j=j_\uparrow = -j_\downarrow$ normalized 
to $j_0 = en_{\rm e} v_{k_F}$ as a function of the ordering wave vector 
$Q$ for $2n_{\rm e}U/(\pi W) = 1$.
The dashed lines indicate an instability regime against phase separation into 
regions with larger and smaller $Q$.}
\label{fig:three}
\end{figure}

Our numerical results are summarized in Figs.~\ref{fig:two} and \ref{fig:three}.
We choose $Q=Q_z$ to be in the $\hat{\bf z}$-direction.
In Fig.~\ref{fig:two} we plot the quasiparticle energies $E^\pm_{\bf k}$ for 
$Q = 0.7 \pi/a$ and $Q = \pi/a$.

The dashed curves represent $\epsilon_{k_z \pm Q/2}$ for spin-up and spin-down
electrons, respectively; note that the bands are degenerate for $k_z a=\pi$ when
the order parameter vanishes. 
The quasiparticle bands $E^\pm_{k_z}$ (solid lines) are split by 
$\sqrt{(\epsilon_{{\bf k} + {\bf Q}/2} -
        \epsilon_{{\bf k} - {\bf Q}/2})^2 + 4 h^2}$, weakening the 
dispersion of both the occupied band and the empty band.
The self-consistently determined values of the order parameter $s_Q$, the
magnetic condensation-energy density $\epsilon_{\rm cond} = \epsilon_0 - 
\epsilon(Q)$ (where $\epsilon_0 = n_{\rm e}W (1/2-1/\pi)$ is the energy 
density of the state with $h=0$), 
and the spin supercurrent density $j = j_{\uparrow} = 
- j_{\downarrow} = $ are plotted as a function of the spiral wave vector
$Q$ in Fig.~\ref{fig:three}.
For the half-filled band case we consider the condensation energy is maximized 
for $Q=\pi/a$. In accord with Eq.~(\ref{eq:curdens}) no spin
supercurrent is present at this
value of $Q$.
For other values of $Q$ the derivative $\partial \epsilon(Q) / \partial Q$ 
is finite and the spin supercurrent is non-zero.
In Fig.~\ref{fig:three} we have used dashed lines in the regions where 
$\partial j(Q)/ \partial Q$ is negative, to emphasize that the spiral
state is not metastable. Energy can always be gained in this regime
without changing the total number of phase windings by introducing
domains with different phase winding rates.
 
This example clearly shows that spiral magnetization configurations do not 
necessarily lead to spin supercurrents;  
instead they occur when the system is driven from its ground
state into a state with a non-optimal spiral wavevector $Q$.  
In the case of a ferromagnet, {\em i.e.}, a system for which the energy
is minimized by $Q=0$, it might\cite{us} be possible to do this by 
driving the system from equilibrium using spin-selective transport currents 
supplied by ferromagnetic electrodes.  This possibility has not yet 
been demonstrated experimentally, however, and may require combinations of material
characteristics and geometries that are difficult to achieve. 
This difficulty motivates trying to find other strategies for realizing persistent
spin currents.  As we discuss in Section IV and V of this paper, in the case of a system 
with a spiral ground state persistent spin currents can be generated simply by 
applying an external magnetic field in the $\hat{\bf x}$-$\hat{\bf y}$ plane.
The in-plane field drives the spiral state to a soliton lattice state which 
{\em does} carry a persistent spin current.
Before turning to this central portion of our paper, 
we first briefly comment on the relationship between our 
discussion of persistent currents in ferromagnets and Anderson's
discussion\cite{anderson} of superconductivity in terms of magnetic order in
an effective spin model.

\section{Analogy with BCS Theory}

To understand the physics behind our results for the dissipationless
spin currents it is instructive to invert Anderson's analogy and
address magnetic properties in terms of superconductivity.
An analogy between Josephson junctions and tunnel junctions
between ferromagnetic metals has been presented recently\cite{bennemann} 
which is in the same spirit as the present discussion.
The spin supercurrents discussed in this paper and the electrical 
supercurrents supported by a Cooper pair condensate in BCS superconductors
appear in a similar way when a particle-hole transformation is performed
in one of the spin subspaces, say for spin down, to convert
particle-particle order into particle-hole order. 
For example, the definition of the order 
parameter in the microscopic model, $s_{\bf Q} = (1/V){\sum_{\bf k}} \langle 
c^\dagger_{{\bf k}+{\bf Q}/2,\uparrow} c_{{\bf k}-{\bf Q}/2,\downarrow} 
\rangle$, is mapped to the order parameter of a superconductor with 
a momentum ${\bf Q}$ pair condensate by this transformation.
A finite value of the order parameter $s_{\bf Q}$ corresponds to the formation
of ``Cooper pairs'', consisting of an electron and a hole.
These pairs carry no electric charge 
so that no charge supercurrent arises from the order.
The formation of pairs with a finite momentum is partly analogous to 
the process leading to 
Fulde-Ferrell-Larkin-Ovchinnikov states\cite{ff} in a superconductor.

We can achieve a qualitative understanding of our numerical results for the
persistent spin current by combining Eq.~(\ref{eq:curdens}),
Eq.~(\ref{eq:qpenergies}) and Eq.~(\ref{eq:meanfieldenergy}), and
ignoring the dependence of the order parameter magnitude on ${\bf
  Q}$. Then the contribution to the current from each wavevector ${\bf
  k}$ in Eq.~\ref{eq:meanfieldenergy} depends only on the velocities of
the electron and the hole in the ``Cooper pair'', ${\bf v}_{\bf k} =
\partial \epsilon_{\bf k} / \partial (\hbar {\bf k})$ and ${\bf v}_{-\bf
  k} = -{\bf v}_{\bf k}$. (These bare velocities can be identified with
the slopes of the dashed lines in Fig.~\ref{fig:two}.) At finite $\bf
Q$, however, the velocity of a ``Cooper pair'' is finite and the sum
over all wavevectors ${\bf k}$ in general produces a result that is not
zero. Thinking in this way we can see directly that the point $Q =
\pi/a$, for which we found in the previous section that the condensation
energy is maximal and the total spin supercurrent is zero, is
special. The reason is that the contribution of each ``Cooper pair'' is
neutralized by that of another pair with reversed wave vectors for both
electron and the hole.

\section{Persistent Spin Currents in Helimagnets}

Two classes of helimagnets have been extensively studied
in the past. The first is comprised of MnSi and related materials that have 
long-period helical ground states because of the Dzyaloshinskii\cite{dzyaloshinskii} 
instability of ferromagnetism in systems without inversion 
symmetry. These systems have been studied extensively very
recently\cite{qcrit} because they provide an example of an
itinerant-electron magnetic system in which the ordering temperature can
conveniently be driven to zero by applying pressure, and because they
appear to show non-Fermi-liquid behavior associated with this nearby
quantum critical point. They also have the potential advantage for the
phenomenon of interest here that the chirality degeneracy, present in
the microscopic model studied in Section II for example, is
lifted\cite{bak2} by broken inversion symmetry. In this case however,
the plane of the spiral magnetic order is fixed only indirectly by
anisotropy in the gradient term in the Landau-Ginzburg energy functional
and the easy plane anisotropy that we require is rather weak.

The second class of materials consists of the 
heavy metals Tb, Dy, and Ho which have helical ground states due
to frustration induced by RKKY interactions between the
rare-earth moments and do have strong easy-plane anisotropy.\cite{elliot,steinitz,plumer}
At present, it appears to us that these are the most promising materials for the realization 
of persistent spin currents. 

In the case of a superconductor, charge supercurrents can be generated 
quite simply by biasing the sample with an external current source.
In the case of an easy-plane ferromagnet, spin supercurrents can be 
generated by biasing the systems with an external spin current.  One possible
scheme for realizing such a bias using four ferromagnetic contacts is 
discussed in Ref.~\onlinecite{us}, but its success depends on avoiding 
spin-flip processes at the interfaces and on having thin-film 
samples with spin-diffusion lengths that are larger than the sample size.  
For systems with a spirally-ordered ground state, however, there is an
easier and more straightforward way of generating persistent spin currents 
which does not depend on achieving spin-current biasing.
Instead, persistent spin currents can be generated in the ground state of the 
system simply by applying a magnetic field in the plane defined by the spiral
order parameter.  
In the continuum model that we study,
the persistent-current state in a helimagnet is a soliton-lattice state 
for which the lattice Bloch wavevector is not a good quantum number.
For this reason we cannot easily describe these states microscopically.
We therefore use a phenomenological model for the following discussion.

\subsection{Magnetic order in the presence of an in-plane magnetic field}

A magnetic field in the $\hat {\mathbf x}$-$\hat {\mathbf y}$ plane
alters the spiral state. At sufficiently large field strengths, it is
clear that the magnetic field will lead to nearly uniform
spin-polarization along the field direction. As we show, for
fields below a critical strength, the spiral state is driven by the
field to a soliton-lattice state, in which the magnetization orientation
slips periodically from the field-direction orientation. 

We assume that due to easy-plane anisotropy the magnetization is in the
$\hat {\mathbf x}$-$\hat {\mathbf y}$ plane and has a spatially constant amplitude.
The orientation variation of the magnetization density 
${\bf m}({\bf r}) = g \mu_{\rm B} \langle {\bf s} (\bf r) \rangle$ 
is parametrized by the angular variable $\vartheta({\bf r})$ via
${\bf m}({\bf r}) = m_0  [\cos(\vartheta({\bf r})) \hat {\bf x} 
- \sin(\vartheta({\bf r})) \hat {\bf y}]$. 
For clarity, we choose $\bf Q^\star$ in $\hat{\bf z}$-direction, 
${\bf Q^\star}=(0,0,Q^\star)$ so that $\vartheta$ only depends on the
$\hat{\bf z}$-position.
However, the following discussion does not depend on this specific choice.
To determine the magnetic order
at $T=0$ we have to minimize the energy density 
\begin{equation}
  \label{eq:PT}
  \epsilon = \int {dz \over L_z} \; 
  \left[ \frac{\rho_{\rm s}}{2} |\partial_z \vartheta({z})- 
    Q^\star|^2 - {\bf B \cdot m}({z}) \right] \, ,
\end{equation}
where $L_z$ is the length of the sample in $\hat{\bf z}$-direction, the
spin stiffness $\rho_{\rm s}$ characterizes the energy cost for spiral
order with wavectors close to the minimal value ${\bf Q^\star}$, and
${\mathbf B}$ is the in-plane magnetic field.

Without loss of generality we choose the inplane magnetic field along the
$\hat{\bf x}$ direction, ${\bf B} = (B,0,0)$.
For this particular choice of directions, $\vartheta$ is the angle between the 
local magnetization direction and the external magnetic field. 
This model for a spiral state in an external field (Eq.~(\ref{eq:PT}))
is equivalent to the Pokrovsky-Talapov (PT) model, reviewed for
example by P.~Bak,\cite{bak} used originally to model
commensurate-incommensurate transitions and more recently to model the
influence of an in-plane field on spontaneous coherence broken symmetry
states in bilayer quantum Hall systems. 
In using this model we assume that the energy density depends on the 
gradient of the orientation angle $\vartheta$ and that, for qualitative 
purposes, we can expand around the value of the gradient that minimizes the 
energy density.  
This chiral model favors a particular sign for $\partial_z \vartheta$ and 
does not account for the symmetry-based expectation that the 
energy density will have minima for $\partial_z \vartheta = \pm Q^*$. 
Hence it accounts only for the effect of a magnetic field on a spiral state 
domain with a particular chirality; we speculate later on persistent spin
currents near boundaries between domains with opposite chirality.
A more general model of the gradient energy that does not favor a 
particular chirality would have the form 
$\int (dz/L_z) (\rho_{\rm s}/2) [-(\partial_z \vartheta)^2/2 +
(\partial_z \vartheta)^4/4Q^{*2}]$ plus a constant.  
Finally, our continuum model ignores the possibility of 
locking to configurations that are commensurate with the 
underlying lattice.  Locking to commensurate configurations has been
studied for helimagnets in fields for both non-chiral\cite{harris}
and chiral\cite{yokoi} models.  Indeed, magnetoelastic coupling is known
to play an important and complex role\cite{plumer} in the response of 
helimagnets to external magnetic fields. Our use of a chiral continuum model
for the following discussion certainly does not capture the full
richness of real helimagnets in external magnetic fields.  Our intention
here is to demonstrate simply the general property that external fields induce
persistent spin currents within each chiral domain, using a model 
that is well understood\cite{bak,hanna}. In the following discussion we 
use well known properties of the classical sine-Gordon model to evaluate 
the persistent spin current within a chiral domain of a helimagnet.

Minimization of the energy with respect to variations in $\vartheta(z)$ 
leads to the sine-Gordon equation
\begin{equation}
  \label{sine-gordon}
  {\partial^2 \vartheta (z) \over \partial z^2} = 
  \frac{1}{\xi^2} \sin(\vartheta(z))
\end{equation}
where $\xi = \sqrt{\rho_{\rm s}/Bm_0}$.
For fields $B$ larger than a critical value $B_c$, the ground state magnetization is uniform
with $\vartheta(z) \equiv 0$.
For $B$ just below $B_c$ the energy can be lowered by incorporating isolated 
$2 \pi$ solitons; $\vartheta_{\rm ss}(z) = 4 \arctan[\exp(\pm (z-z_0)/\xi)]$
for a soliton centered at $z_0$.
At $B_c$, the uniform state has the same energy as a state which
accommodates a single soliton.  By comparing energies we find that
in terms of the notation defined above
\begin{equation}
  B_c = \frac{\pi^2}{16} \frac{\rho_{\rm s}}{m_0}{Q^\star}^2 \, .
\end{equation}
For $B<B_c$ the magnetic state can be described as a soliton lattice
(SL) with period $a$. The associated wave vector $Q =2\pi/a$ varies from
$Q=0$ at $B=B_c$ to $Q=Q^\star$ at $B=0$.  In the latter limit the
soliton-lattice state approaches the spiral magnetic order state,
discussed from a microscopic point of view in the first part of the
paper. The total phase change along the sample is given by $QL_z$. 

\begin{figure}
\begin{center}
\includegraphics[width=0.9\columnwidth]{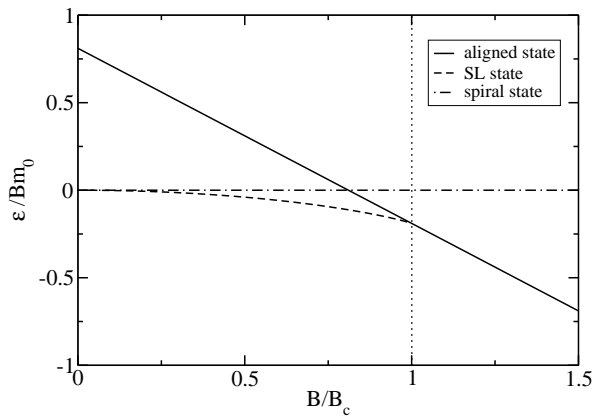}
\caption{\label{fig:four} Energy versus in-plane magnetic field
for our spiral state model.  The soliton-lattice state minimizes
the model energy for $B < B_c$, while the state in which the 
magnetization is aligned with field minimizes the energy for 
$B > B_c$.  The soliton lattice state energy approaches the spiral state 
energy as $B \to 0$.  The energies are normalized to $B_c m_0$.}
\end{center}
\end{figure}

The results collected in the following paragraph are adapted from
Ref.~\onlinecite{hanna} and earlier work cited therein.
Minimization of the model's energy in the soliton lattice state gives,
$\cos(\vartheta(z)/2) = - {\mathrm{sn}}[(z-z_0)/\eta\xi,\eta]$
where $\mathrm{sn}$ denotes the sine-amplitude Jacobian elliptic
function,\cite{gradshteyn} and $\eta$ is a constant which depends on the
strength of the magnetic field $B$;
\begin{equation}
  \frac{B}{B_c}=\left(\frac{\eta}{E(\eta)}\right)^2 \, ,
\end{equation}
where $E(\eta)\equiv \int_0^{\pi/2} d \beta  \sqrt{1- \eta ^2 \sin ^2 \beta }$
denotes the complete elliptic integral of the second kind.\cite{gradshteyn}
From that we see that the limit $\eta = 1$ applies at 
the SL state with $B=B_c$, and that $\eta = 0$ describes the opposite 
situation of a vanishing in-plane magnetic field $B = 0$.
The wave vector $Q$ of the SL is given by 
\begin{equation}
  \frac{Q}{Q^\star} = \frac{(\pi/2)^2}{K(\eta) E(\eta)} \, ,
\end{equation}
where $K(\eta)\equiv \int_0^{\pi/2} d \beta /\sqrt{1- \eta^2 \sin^2 \beta}$
denotes the complete elliptic integral of the first kind.\cite{gradshteyn}
We note that $Q=0$ for $B=B_c$, and $Q=Q^\star$ for $B=0$.
Finally, the expression for the energy density of the SL, which we relate
to persistent spin currents below, is 
\begin{equation}
  \label{eq:endens}
  \epsilon = \frac{\rho_{\rm s}}{2} {Q^\star}^2-Bm_0 \left(
    \frac{2}{\eta^2}-1 \right) \, .
\end{equation} 
The SL energy density is plotted as a dashed line in Fig.~(\ref{fig:four}). 
For $B<B_c$ the SL is energetically favored over a uniform state, whose
energy is $(1/2) \rho_{\rm s} {Q^\star}^2-Bm_0$.
For $B\neq 0$ the SL lattice is always favored over the spiral state (with wave 
vector $Q^\star$).
The energy of the latter state is the zero of energy of our model.

A cautionary note is appropriate at this point.
We do not expect that the expansion of the gradient energy around the
spiral wavevector that appears in our model will be realistic for
typical spiral magnets all the way down to the $Q=0$ state that is
favored by the in-plane field.  It follows that our model
is most realistic physically in the $B < B_c$ region. 

\subsection{Spin supercurrent}

\begin{figure}
\begin{center}
\includegraphics[width=0.9\columnwidth]{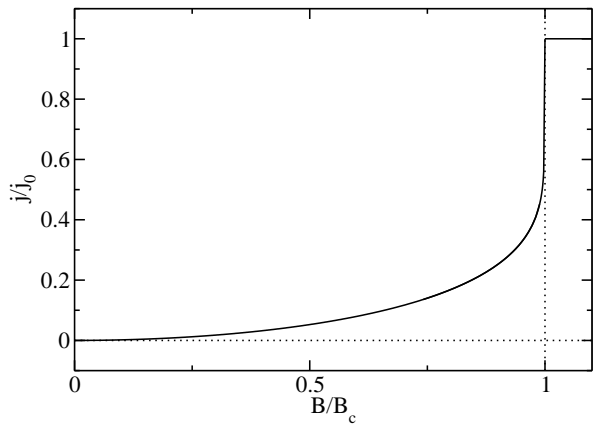}
\caption{\label{fig:five}
  Spin supercurrent density $j\equiv j_\uparrow$ induced by an in-plane 
  magnetic field.
  In the spirally ordered state at $B = 0$ there is no current and for
  $B>B_c$ the current is constant.}
\end{center}
\end{figure}

We now demonstrate how the in-plane magnetic field gives rise to a persistent 
spin current.  To derive an expression for the spin current 
density we introduce a 
spin-dependent, spatially constant vector potential 
${\mathbf A_\sigma} = A_\sigma \hat {\mathbf z}$. 
The energy density of the system then reads
\begin{eqnarray}
\label{eqn:current_deriv}
  \epsilon =  \frac{1}{2}\sum_\sigma && \!\!\!\! \int {dz\over L_z} \; 
  \left[
  {\rho_{\rm s} \over 2} \left({\partial \vartheta \over \partial z} 
      - Q^\star + \sigma \frac{2e}{\hbar c}A_\sigma \right)^2 
  \right. 
\nonumber \\
    && \left.  
      - B m_0 \cos\left(\vartheta + \frac{e(A_\uparrow - A_\downarrow)}
        {\hbar c} z \right)\right] \, .
\end{eqnarray}
Note that here we have explicitly used the coincidence of $\vartheta$
with the angle between the magnetization direction and the magnetic
field; for a different orientation of the magnetic field in the 
$\hat{\bf x}$-$\hat{\bf y}$ plane, Eq.~(\ref{eqn:current_deriv}) adopts a 
different form.
The spin-dependent current density follows from the derivative
$j_\uparrow = c \partial \epsilon(A_\uparrow,A_\downarrow) / \partial 
A_\uparrow$ at $A_\uparrow = A_\downarrow = 0$.
Making use of the relation $\int dz \sin\left(\vartheta(z)\right) = 0$, that
holds for all soliton lattices with arbitrary wave vector, we find that
\begin{equation}
  j_\uparrow = 
  \frac{e}{\hbar} \rho_{\rm s} \left( Q - Q^\star \right) 
  = - j_\downarrow \, .
\end{equation}
At $B=0$ we have a spiral state with $Q=Q^\star$, and the spin current 
vanishes.
With increasing strength of the $B$ field the wave vector $Q$ decreases and
a finite spin supercurrent arises.
At $B=B_c$ the spin current $j_\uparrow$ reaches its maximal value
$j_0\equiv - (e/\hbar) \rho_{\rm s} Q^\star$.
The limit of small magnetic fields, $B \ll B_c$, is described by
\begin{equation}
  \frac{j_\uparrow}{j_0} \approx \frac{\pi^4}{512}
  \left(\frac{B}{B_c}\right)^2 \, .
\end{equation}
The spin current density as a function of the in-plane magnetic field is
shown in Fig.~\ref{fig:five}.

In conclusion, the SL state of a helimagnet carries a finite persistent
spin current, without any electrical current, when a magnetic field
is applied in the plane defined by the spiral order.

\section{Discussion}

We have shown in this paper that persistent spin currents can be induced
in helimagnets by applying a magnetic field in the plane defined by the 
spiral order.  It is interesting to speculate on the possibility of exploiting 
this effect for new types of spintronic devices.
Current spintronic devices make use of magnetotransport effects like 
anisotropic magnetoresistance, giant magnetoresistance,\cite{gmr} and tunnel
magnetoresistance,\cite{tmr}
all of which follow from the dependence of quasiparticle transport on the 
collective magnetic state. 
The effect we have discussed here is distinctly different 
in that it is an equilibrium spin current that is carried collectively
rather than by quasiparticles.

\begin{figure}
\begin{center}
\includegraphics[width=0.9\columnwidth]{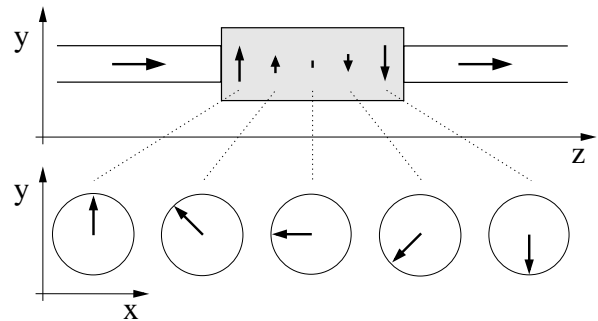}
\caption{\label{fig:six}
Sketch of the setup discussed in the text. The direction of the
magnetization is indicated by the arrows. In the wires it is parallel to the $\hat{\bf z}$-axes along
the transport direction, while in the helimagnet (shaded grey) it is
circulating in the $\hat{\bf x}$-$\hat{\bf y}$ plane as shown in the lower part of the sketch.}
\end{center}
\end{figure}

To illustrate the possibility of realizing interesting magnetotransport effects
based on field-induced persistent currents in helimagnets, consider the 
case illustrated in Fig.~\ref{fig:six}.  We consider current flow along a ferromagnetic
wire which contains a helimagnet.  The magnetization direction in the ferromagnets
is assumed to be aligned along the direction of the wire by magnetostatic energy,
and the electrical current in the ferromagnet will be spin-polarized along 
the direction of the magnetization.  We also assume that the helimagnet is grown so 
that it has its wavevector along the wire direction and the spiral magnetization
is in the perpendicular plane; this would be the case for example for rare earth 
metal single-crystal helimagnets with their hcp c-axis aligned with the wire.
The quasiparticles of the helimagnet would then tend to have their spins 
aligned in the plane perpendicular to the wire axis, suppressing their ability 
to carry currents that are spin-polarized along the wire axis, and increasing the 
resistance of the overall system.  In the presence of a field perpendicular
to the wire axis, however, the helimagnet would have a persistent spin current 
polarized along the wire axis, eliminating the need to carry the spin current
with quasiparticles.  We predict that because of the persistent spin currents, 
an in-plane field could therefore alter the resistance of this system. 
Since the persistent spin current has a sign that depends on the chirality of 
the spiral, the sign of the resistance depends on this quantity.  
The effect should be strongest in samples with a single chiral domain.  

This same effect could also be of interest in a similar setup where the
helimagnet is connected to normal metallic leads. Then the spin current
in the helimagnet results in a different resistance for the electron
spins entering from the leads. In this fashion the helimagnet could act
as a tunable spin filter which is controlled by an external magnetic field.

It may also be possible to realize spin torque\cite{torque} effects in helimagnets
that are distinct from those which occur in ferromagnets. Spin torque
occurs when the spin polarization of a quasiparticle current changes
with position by altering the spin distribution of current-carrying
states, transferring the spin carried by the quasiparticle to the
collective magnetic coordinate. The field-induced persistent spin
current in a helimagnet will have a strong spatial dependence, a sign
change in fact, near a boundary surface between domains with opposite
chirality. The divergence of this spin current makes a contribution to
the time-dependent spin polarization that is cancelled by the torque due
to the external magnetic field when the domain wall is in
equilibrium. When the domain wall is not in equilibrium, for example by
being displaced from a pinning center, these two contributions will not
cancel and the net torque will lead to a time-dependent magnetization
orientation. Conversely, the equilibrium position of a pinned domain wall will
be sensitive to the presence of persistent currents and will therefore
be altered by an in-plane field. 

In conclusion, we have studied the relation between spiral magnetic order
and dissipationless spin transport.
We have demonstrated the possibility of dissipationless spin currents in states
with equilibrium quasiparticle populations. 
These spin currents are always associated with spiral magnetic order.
On the other hand, the existence of spiral magnetic order is, in general, not 
sufficient to guarantee collective spin transport.
In particular helimagnets, which have spiral ground states,
do not support persistent spin currents.
However, persistent spin currents can be induced in these systems 
simply by applying an external magnetic field.  

We acknowledge helpful discussions with Martin B{\o}nsager, Herbert
Capellmann, Jan Martinek, Herbert Schoeller, Michael Steinitz and Leo
Radzihovsky. The work is part of the CFN which is supported by the
DFG. We acknowledge funding by the European Training Network
RTN2-2001-00440 and by the DFG-Emmy-Noether-program KO1987/2. Work at 
the University of Texas was supported by the Welch Foundation and by the
DOE under grant DE-FG03-02ER45958.

\appendix
\section{Proof of Eq.~(2)}
\label{proof}

In order to prove that the spin-current density is related to the 
derivative of the total energy with respect to the wave vector $\bf Q$ as
given in Eq.~(\ref{eq:curdens}) we extend the Hartree-Fock Hamiltonian 
Eq.~(\ref{eq:mft}) by formally introducing an additional parameter 
$\bf \tilde Q$ via the replacement $\epsilon_{{\bf k}+{\bf Q}/2} \rightarrow 
\epsilon_{{\bf k}+({\bf Q}+{\bf \tilde Q})/2}$ and
$\epsilon_{{\bf k}-{\bf Q}/2} \rightarrow 
\epsilon_{{\bf k}-({\bf Q}+{\bf \tilde Q})/2}$.
Our original Hamiltonian Eq.~(\ref{eq:mft}) corresponds to 
${\cal H}^{\rm HF}({\bf Q},{\bf \tilde Q})|_{{\bf \tilde Q}=0}$.
The quasiparticle ground state for given ${\bf Q}$ and ${\bf \tilde Q}$ is an 
eigenstate and, therefore, the Hellmann-Feynman theorem applies,
\begin{equation}
  {\partial \langle {\cal H}^{\rm MF} ({\bf Q},{\bf \tilde Q})\rangle \over
   \partial {\bf \tilde Q}} \bigg|_{{\bf \tilde Q}=0} =
  \left\langle {\partial {\cal H}^{\rm MF} ({\bf Q},{\bf \tilde Q}) \over
   \partial {\bf \tilde Q}} \right\rangle \bigg|_{{\bf \tilde Q}=0} 
\label{eq:Hellmann}
\end{equation}
By comparing the r.h.s with Eq.~(\ref{eq:defj}) we find that 
Eq.~(\ref{eq:Hellmann}) equals to 
$V(\hbar / 2e) ({\bf j}_\uparrow - {\bf j}_\downarrow)$.
To evaluate the l.h.s of Eq.~(\ref{eq:Hellmann}) we observe that the 
quasiparticle energies $E^\pm_{\bf k}$ for 
${\cal H}^{\rm MF} ({\bf Q},{\bf \tilde Q})$ as well as the value of $h$ 
depend on ${\bf Q}$ and ${\bf \tilde Q}$ only in the combination 
${\bf Q}+{\bf \tilde Q}$.
As a consequence, we can replace $\partial/\partial {\bf \tilde Q}$ by 
$\partial/\partial {\bf Q}$.
Together with ${\bf j}_\uparrow = - {\bf j}_\downarrow$, this immediately
proves Eq.~(\ref{eq:curdens}).

\end{document}